\documentclass{owrart}

\usepackage{xcolor}

\newcommand{\bu}{\mathbf{u}}

\newcommand{\br}{\mathbf{r}}

\begin{document}


\begin{talk}[Damien Fournier, Thorsten Hohage]{Laurent Gizon}
{Problems in computational helioseismology}
{Gizon, Laurent}

\noindent

The Sun supports acoustic oscillations continuously excited by near-surface turbulent convection.  {\it Global helioseismology} consists of inverting the measured frequencies of the normal modes of oscillation to infer the sound speed and rotation as a function of radius and unsigned latitude \cite{Basu2016}. 
 Techniques of {\it local helioseismology} based on correlations of  the wave field at the surface are being developed to infer the structure and dynamics of the Sun in three dimensions  \cite{Gizon2010, Hanasoge2016}. 

{\bf Forward problem.} Time-distance helioseismology \cite{Duvall} is a particular technique of local helioseismology,  analogous to geophysical seismic interferometry. Ignoring terms that involve gravity, the oscillations at position $\br$ and frequency $\omega$ can be described by a scalar field $\psi(\br, \omega)$, which solves the acoustic wave equation  \cite{Gizon2017}
\begin{equation}
L_{\br,\omega} [\psi] : = 
-(\omega^2 + 2i \omega\gamma)\psi  -2i \omega \bu\cdot\nabla_{\br}  \psi - c \nabla_\br\cdot\left(\frac{1}{\rho } \nabla_\br ( \rho c \psi ) \right) = s(\br,\omega)   ,
\end{equation}
where $\gamma(\br,\omega)$ is attenuation and the steady background medium is represented by density $\rho(\br)$, sound speed $c(\br)$, and flow $\bu(\br)$. Waves are excited by a stationary random process (granulation) represented by the function $s(\br,\omega)$.
The above equation is supplemented by a radiative boundary condition  \cite{Barucq2017}.
The basic input data  in time-distance helioseismology is the covariance function $C(\br',\br,\omega)= \psi^*(\br', \omega) \psi(\br, \omega)$ between two points on the solar surface. Under the assumption that  sources are spatially uncorrelated and of the form $\mathbb{E}[s^*(\br',\omega) s(\br,\omega)] = \delta(\br-\br')   P(\omega) \gamma(\br,\omega)  / \rho(\br) $ we have (to within a surface term)  
\begin{equation}
 C (\br',\br, \omega)  = \frac{P(\omega)}{4i\omega} \left[  G (\br, \br', \omega)  - G^\dagger (\br, \br',\omega) \right]  +  \text{noise}  , 
 \label{eq:CGG}
 \end{equation}
where  $   L_{\br,\omega}[ G(\br,\br',\omega) ] =  \delta(\br-\br')/ \rho (\br) $ and 
$G^\dagger = G^*(\bu\rightarrow -\bu)$ is obtained by switching the sign of $\bu$ and taking the complex conjugate. 
The linear forward problem consists in computing the perturbations to the covariance function caused by infinitesimally small  perturbations in the background medium. Combining the first Born approximation \cite{Gizon2002, Boening2016}
and Eq.~(\ref{eq:CGG}), Gizon et al.  \cite{Gizon2017} expressed sensitivity kernels in terms of only four Green's functions in the reference medium,  computed using the finite-element code Montjoie \cite{Montjoie}.

{\bf Inverse problem.} 
The  inverse problem consists of reconstructing $\gamma(\br,\omega)$, $c(\br)$, $\rho(\br)$, and $\bu(\br)$ in the interior, starting from a reference solar model. This requires knowledge of the noise covariance matrix \cite{Gizon2004,Fournier2014}.
Linear inversions are traditionally performed using Tikhonov regularization \cite{Kosovichev1996} or the
method of approximate inverse (called optimally localized averaging, see ref.~\cite{Pijpers1992}). 
Under the assumption of local horizontal translation invariance of the sensitivity kernels, multichannel inversions in Fourier space enable to solve problems that would otherwise require too much computer memory \cite{Jackiewicz2012}. Minimax estimators have been computed for this problem using the Pinsker method \cite{Pinsker1980, Fournier2016}.

The non-linear inverse problem of time-distance helioseismology (finite pertubations to the medium) has not been studied in full detail yet. Future studies should build on existing theoretical uniqueness results, in particular on the  Novikov-Agaltsov reconstruction algorithm \cite{Agaltsov2016}, which combines measurements of $G$ at several frequencies (see table below).  For measurements of $C$ instead of $G$, we have conducted numerical experiments to determine the number of frequencies required to reconstruct $\rho$ and $c$. For realistic noise levels, more frequencies will be needed to obtain useful reconstructions.
\begin{table}[h]
\label{table:GFH}
\caption{Number of frequencies needed for reconstruction} 
\begin{tabular}{c | c c c c} 
\hline\hline 
  & \multicolumn{2}{c}{Observable: $G$} &
\multicolumn{2}{c}{Observable: $C$}
     \\ [0.5ex] %
 & theory  & experiment  & theory  & experiment \\ [0.5ex] %
 \hline 
$c$ & 1 (ref. \cite{Novikov1988})& 1 (ref. \cite{Nachman1988})& ? & 2 (this work)\\ 
$\rho$ &1 (ref. \cite{Novikov1988})& 1 (ref. \cite{Nachman1988})& ? & 2 (this work)\\
$c$, $\rho$  & 2 (ref. \cite{Novikov1988})& 2 (ref. \cite{Nachman1988})& ? & 4 (this work)\\
$\bu$  &2 (ref. \cite{Agaltsov2016})& 2 (ref.  \cite{Zotov2017}) & ? & 2 (ref.  \cite{Zotov2017})\\
$c$, $\rho$, $\gamma$, $\bu$  &3 (ref. \cite{Agaltsov2016})& $\ge 3$ (ref. \cite{Shurup-priv})  & ? & ? \\
\hline 
\end{tabular}
\end{table}

{\bf Outstanding problems.}  
Further advances in local helioseismology will require improved forward solvers for vector MHD wave equations (see refs.~\cite{Cameron2008, Hanasoge2010}) and homogenized wave equations \cite{Hanasoge2013}, as well as improved inversion methods that minimize the number of forward solves \cite{Hanasoge2011,ref2}. A major challenge in local helioseismology is the very large size of the input dataset, e.g. $\sim 10^{12}$ pairs of points times $\sim 10^2$ frequencies in time-distance helioseismology. As a result, it is important to either select or average the input data before inverting them. One interesting averaging scheme that deserves further attention is helioseismic holography \cite{Lindsey2000a, Lindsey2000b, Skartlien2002}, which uses  Green's second identity to image scatterers in the Sun, as in Porter-Bojarski holography \cite{Porter1982,Devaney1985}.

\end{talk}

\end{document}